\documentclass[10pt,a4paper,twocolumn,aps,superscriptaddress,Xlinenumbers]{revtex4-1}
\usepackage[utf8]{inputenc}
\usepackage{color}
\usepackage{float}
\usepackage{amsmath}
\usepackage{amssymb}
\usepackage{graphicx}
\usepackage[pdfusetitle,
 bookmarks=true,bookmarksnumbered=false,bookmarksopen=false,
 breaklinks=false,pdfborder={0 0 1},backref=false,colorlinks=true]
 {hyperref}
\hypersetup{
 pdfborderstyle=,citecolor=blue}

\makeatletter

\pdfpageheight\paperheight
\pdfpagewidth\paperwidth

\graphicspath{{./}{./figs/}}

\makeatother

\begin{document}
\title{Time Crystals in Actively Mode-locked Lasers }
\author{Ruiling Weng}
\affiliation{Departament de Física and IAC$^{3}$, Universitat de les Illes Balears,
Campus UIB 07122 Mallorca, Spain }
\author{Elias R. Koch}
\affiliation{Institute for Theoretical Physics, University of Münster, Wilhelm-Klemm-Str9,
48149 Münster, Germany }
\author{Jesús Yelo-Sarrión}
\affiliation{Departament de Física and IAC$^{3}$, Universitat de les Illes Balears,
Campus UIB 07122 Mallorca, Spain }
\author{Josep Batle}
\affiliation{Departament de Física and IAC$^{3}$, Universitat de les Illes Balears,
Campus UIB 07122 Mallorca, Spain }
\author{Neil G. R. Broderick}
\affiliation{Dodd Walls Centre for Photonic and Quantum Technologies, Department
of Physics, University of Auckland, Private Bag 92019, Auckland, New
Zealand}
\author{Julien Javaloyes}
\affiliation{Departament de Física and IAC$^{3}$, Universitat de les Illes Balears,
Campus UIB 07122 Mallorca, Spain }
\author{Svetlana V. Gurevich}
\affiliation{Departament de Física and IAC$^{3}$, Universitat de les Illes Balears,
Campus UIB 07122 Mallorca, Spain }
\affiliation{Institute for Theoretical Physics, University of Münster, Wilhelm-Klemm-Str9,
48149 Münster, Germany }
\affiliation{Center for Data Science and Complexity (CDSC),, University of Münster,
Corrensstraße 2, 48149 Münster, Germany }
\begin{abstract}
We report the first experimental observation of discrete time-crystal
phases and crystallites in an actively mode-locked semiconductor laser.
By tuning either the bias current or the modulation frequency, the
system undergoes a spontaneous symmetry-breaking transition from the
harmonically mode-locked state towards robust, highly coherent time-crystal
states that persist indefinitely. Two equivalent time-crystal configurations,
shifted by one driving period, can coexist as domains separated by
sharp, long-lived boundaries analogous to domain walls. The phenomenon
is quantitatively reproduced by a time-delayed model adapted to the
large gain and losses of semiconductor systems. Our findings demonstrate
that mode-locked semiconductor lasers offer a readily accessible platform
to explore and control non-equilibrium phases of light, enabling practical
implementations of time-crystal physics in photonic systems.
\end{abstract}
\maketitle
Breaking continuous or discrete symmetries is a central topic in modern
physics, underlying a wide range of ordered phases of matter. A particularly
intriguing example is the breaking of \textit{time}-translation symmetry,
leading to so-called \textit{time crystals} (TCs). First proposed
by F.~Wilczek in 2012 as quantum systems whose ground state exhibits
temporal periodicity~\citep{W-PRL-12}, the original concept was
soon shown to be unfeasible in isolated configurations~\citep{B-PRL-13,WO-PRL-15}.
This prompted a broader framework encompassing \textit{dissipative}
and \textit{non-equilibrium} systems~\citep{S-PRA-15,WO-PRL-15,EBN-PRL-16,KLM-PRL-16,GHU-PRL-18,BTJ-NC-19},
where robust \textit{discrete} TCs can emerge in periodically driven
platforms. These phases are characterized by a subharmonic response
to the drive~\citep{SZ-RPP-18,S-TC-2020}, producing a ``crypto-equilibrium''\citep{YN-PT-18},
\emph{i.e.} a state that appears stationary under stroboscopic observation,
yet oscillates at a period that is an integer multiple of the driving
period. Analogously to spatial crystals, TCs display long-range order,
robustness, and a distinct phase transition from which the time-periodic
order emerges spontaneously. While discrete TCs have been realized
in trapped atomic ions~\citep{ZHK-N-17}, driven disordered spin
ensembles in diamond~\citep{CCL-N-17}, and superconducting qubit
systems~\citep{FR-SA-22,MIQ-N-22}, photonic demonstrations remain
scarce. In cavity-assisted cold-atom experiments~\citep{KKP-PRL-21},
TC phases emerge spontaneously but can only be sustained for short
observation times due to the complexity of the setup. Optical Kerr
cavities have been proposed as platforms for dissipative TCs~\citep{TMM-NAC-22},
but the TC regimes coexist with numerous other soliton arrangements.
Synchronized intensity dropouts in semiconductor lasers with time-delayed
feedback have been interpreted as TCs~\citep{TAM-SR-22}. Mode-locked
lasers ---despite being many-body photonic systems due to the large
number of interacting modes--- have traditionally been regarded as
unlikely hosts for genuine TC phases, owing to the presumed absence
of long-range temporal order and crypto-equilibrium~\citep{YN-PT-18}. 

\begin{figure}[b]
\includegraphics[viewport=0bp 50bp 558bp 498bp,clip,width=1\columnwidth]{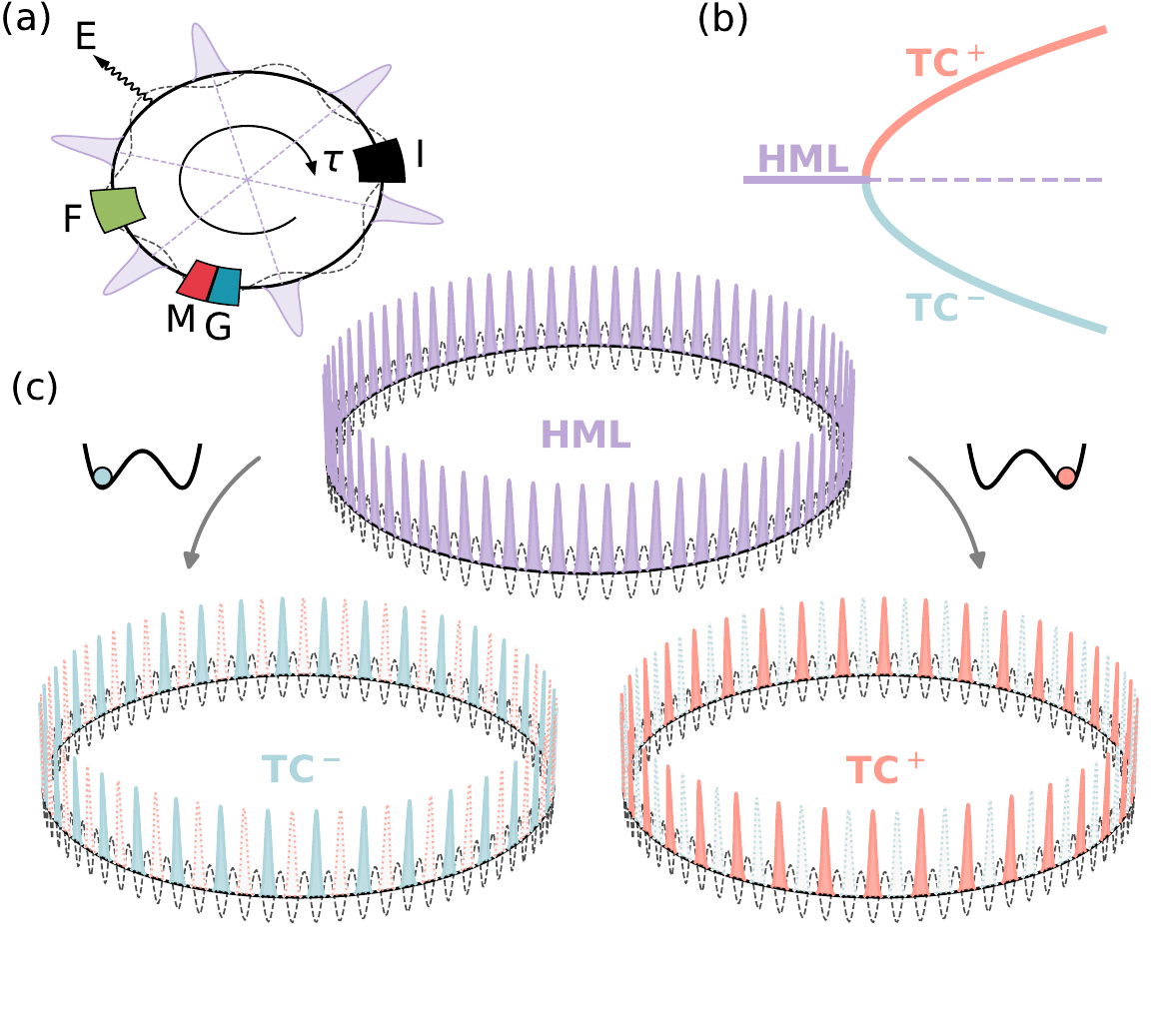}
\caption{(a) Schematic of a unidirectional actively mode-locked ring cavity
operated in the HML regime. (G): gain section; (M): Intensity modulator;
(F): bandpass filter; (I): Optical insulator; (b) Splitting of the
HML state into the two bistable TC phases $\mathrm{TC}^{\pm}$, shown
in (c) together with the HML state from which they emerge.}
\label{fig:1}
\end{figure}

\begin{figure*}[t]
\centering{}\includegraphics[clip,width=2\columnwidth]{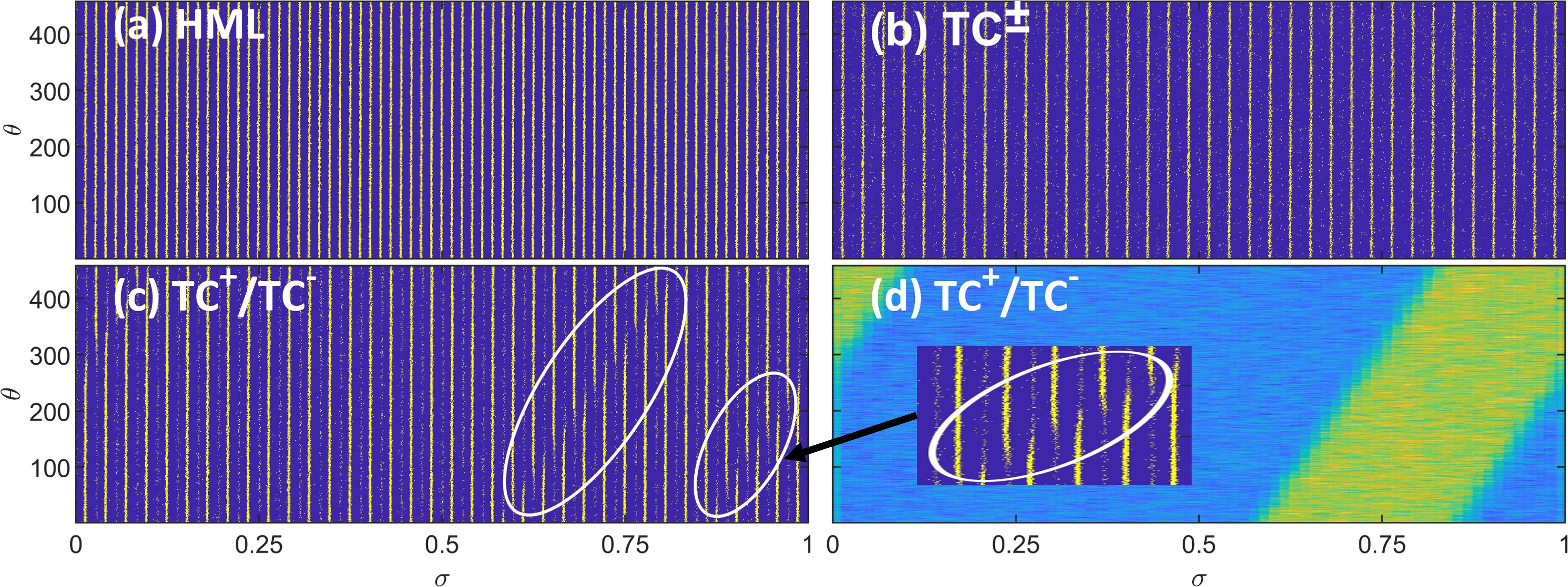}
\caption{(a-d) Bi-dimensional diagrams showing the different observable states.
The horizontal axis represents the (normalized) pulse position within
the round-trip while the vertical axis denotes the number of round-trip.
(a) Regular HML regime with $N=72$ pulses. (b) TC$^{\pm}$ state
with one pulse every two modulation cycles, leading to $N/2$ pulses.
(c) Coexistence between the TC$^{+}$ and TC$^{-}$ states. (d) Processed
data showing the two TC$^{\pm}$ states as constant levels $y_{\pm}=\pm1$.}
\label{fig:2}
\end{figure*}

Here we challenge this view and report the first experimental observation
of a spontaneously emerging discrete TC in an actively mode-locked
(AML) semiconductor laser operating at room temperature. This TC phase
appears when the bias current or the modulation frequency is tuned
near a harmonic mode-locking (HML) resonance, replacing the conventional
HML state without requiring complex experimental conditions. The TC
states we observe are highly coherent and robust, persisting indefinitely.
They occur in two equivalent configurations (TC$^{+}$ and TC$^{-}$),
shifted by one period of the driving modulation, and we experimentally
verify their bistability and complete equivalence. Furthermore, we
reveal composite states ---or ``crystallites''--- where domains
of each TC configuration coexist, separated by sharp, long-lived transitions
that behave as domain walls. These results are well reproduced by
a time-delayed theoretical model adapted to the large gain and loss
typical of semiconductor AML systems~\citep{KGJ-OL-24}. Our findings
position AML semiconductor lasers as compact, tunable, and experimentally
accessible platforms for exploring non-equilibrium phases of light.
By enabling long-lived photonic time-crystal behavior in a robust
and controllable setting, this work bridges the gap between fundamental
concepts of discrete time-translation symmetry breaking and their
potential applications in precision metrology, frequency comb engineering,
and photonic information processing.

Despite their inception dating back to 1964 \citep{HFP-APL-64}, mode-locked
lasers remain an active subject of research due to their rich and
complex dynamics and are crucial for a wide range of applications
ranging from medicine to laser metrology \citep{WSY_APLP_16,HMCD_JLT_00,HZU_IEE_01,HUH_PRL_00,UHH_N_02,K_Nt_03,AJ-BOOK-17}.
They are extensively used as highly coherent light sources~\citep{ST_OLT_15}
to produce ultrashort pulses and frequency combs~\citep{RHWC_NC_16,YZC_OE_20}.
Mode-locking refers to the coherent superposition of numerous lasing
modes within a cavity, resulting in a sequence of identical optical
pulses. Active mode-locking promotes the pulsed operation and the
locking of modes using either an electro-optic or acousto-optic modulator~\citep{TK_IEE_09,CDDC_OL_09,KKL_PhRes_17,KKWK_OL_94,PYZ_OC_02,ZHM_OC_05}
to gate photons within the laser cavity. A key advantage of AML is
its electrical tunability, which enables controlling the pulse shape
and repetition rate \citep{PGG-NAC-20,BWX_AFM_18}. Driving the modulator
with a periodicity $T$ being an integer fraction of the cavity round-trip
$\tau$, \emph{i.e.} $T=\tau/N$ with $N\in\mathbb{N}$, leads to
the emission of $N$ equidistant pulses per round-trip, and is usually
called harmonic mode-locking regime. High harmonic numbers lead to
a system with many degrees of freedom and possibly thousands of individual
pulses circulating in the cavity. This fact, in combination with slow
dynamics typical of gain media such as Ytterbium, Neodymium or Erbium,
may lead to unstable regimes and dropout phenomena \citep{ZPH-JQE-05,ZH-JLT-06,LSB-OECC-07}
although a variety of techniques have been developed to regularize
pulse trains and improves its coherence \citep{YKN-APL-92,HM-OL-93,SS-EL-93,DHI-OL-94,CD-OL-96,TGK-OL-00}
. In our experiment the semiconductor fast gain promotes strong pulse
interaction that regularizes the dynamics \citep{KCB-JQE-98,CJM-PRA-16}.

We depict in Fig.~\ref{fig:1}(a) a sketch of the experimental setup.
An optical insulator ensures unidirectional light propagation while
a linear band-pass filter controls the pulse-width and central frequency.
The cavity length is $L\sim11\,$m which corresponds to a round-trip
of $\tau\sim54.5\ $ns. The gain is provided by a fibered semiconductor
booster amplifier operating around $\lambda=1550$~nm. The mode-locker
element is an intra-cavity Mach-Zehnder electro-optical modulator
(MZM) driven by a radio-frequency (RF) signal generator. The cavity
is composed of polarization maintaining fiber and the laser only emits
on the slow axis of the cavity due to the polarization dependent gain
of the amplifier, see Appendix 1 for more details. Upon tuning the
bias current or the frequency of the RF generator around the $N$-th
HML resonance tongue~\citep{KGJ-OL-24}, we observe that the two
equivalent configurations of the TC, relatively shifted by one period
of the driving modulation and denoted $\mathrm{TC}^{\pm}$, emerge
from the HML$_{N}$ solution, see Fig.~\ref{fig:1}(b,c). Consequently,
the $\mathrm{TC}^{\pm}$ states spontaneously break the discrete time-translation
symmetry imposed by the modulator. 

We present in Fig.~\ref{fig:2}(a) our experimental observations
in the case of the HML regime with $N=72$ pulses per round-trip.
There, we represent the intensity output over $450$ round-trips as
bi-dimensional maps \citep{AGL-PRA-92}, where $\sigma$ is the time
within the round-trip and $\theta$ the round-trip number. The same
solution is depicted schematically in Fig.~\ref{fig:1}(b,c) in purple.
Tuning the bias current or the frequency of the RF generator around
the $N$-th HML resonance tongue allows to see the spontaneous emergence
of the TC states, see Fig.~\ref{fig:2}(b). There, the characteristic
sub-harmonic response of TCs appears at half of the drive frequency
as every other pulse disappear from the potential minima created by
the intra-cavity modulator. The two resulting configurations, denoted
as TC$^{\pm}$ and schematically shown in Fig.~\ref{fig:1}(b,c),
cannot be distinguished in the experiment when they appear. However
we were able to observe situations where both the TC$^{+}$ and TC$^{-}$
states coexist and are separated by sharp transitions, or defects,
as shown in Fig.~\ref{fig:2}(c). There, the ellipses mark the transitions
between the two states. Such states can persist for several minutes
showing the complete equivalence between the two TC states. Any difference
between these states would lead to slow coarsening towards the most
stable TC phase that is favored by the system \citep{GMZ-EPL-12,JAH-PRL-15}.
For better visibility, we map the two TC states to two constant levels
by multiplying the time trace with an alternating function, \emph{e.g.}
a cosine, at half the frequency of the modulation, so that pulses
in even and odd slots gain an alternating sign. Then, an averaging
filter was applied resulting in the bi-dimensional diagram presented
in Fig.~\ref{fig:2}(d), stressing the domain walls analogy with
co-existing TC states and facilitating their experimental identification
in real time. Situations with more than two domain walls (up to 6)
were also experimentally observed, although the domain walls always
appear in pairs for $N=72$. 

\begin{figure}
\includegraphics[width=1\columnwidth]{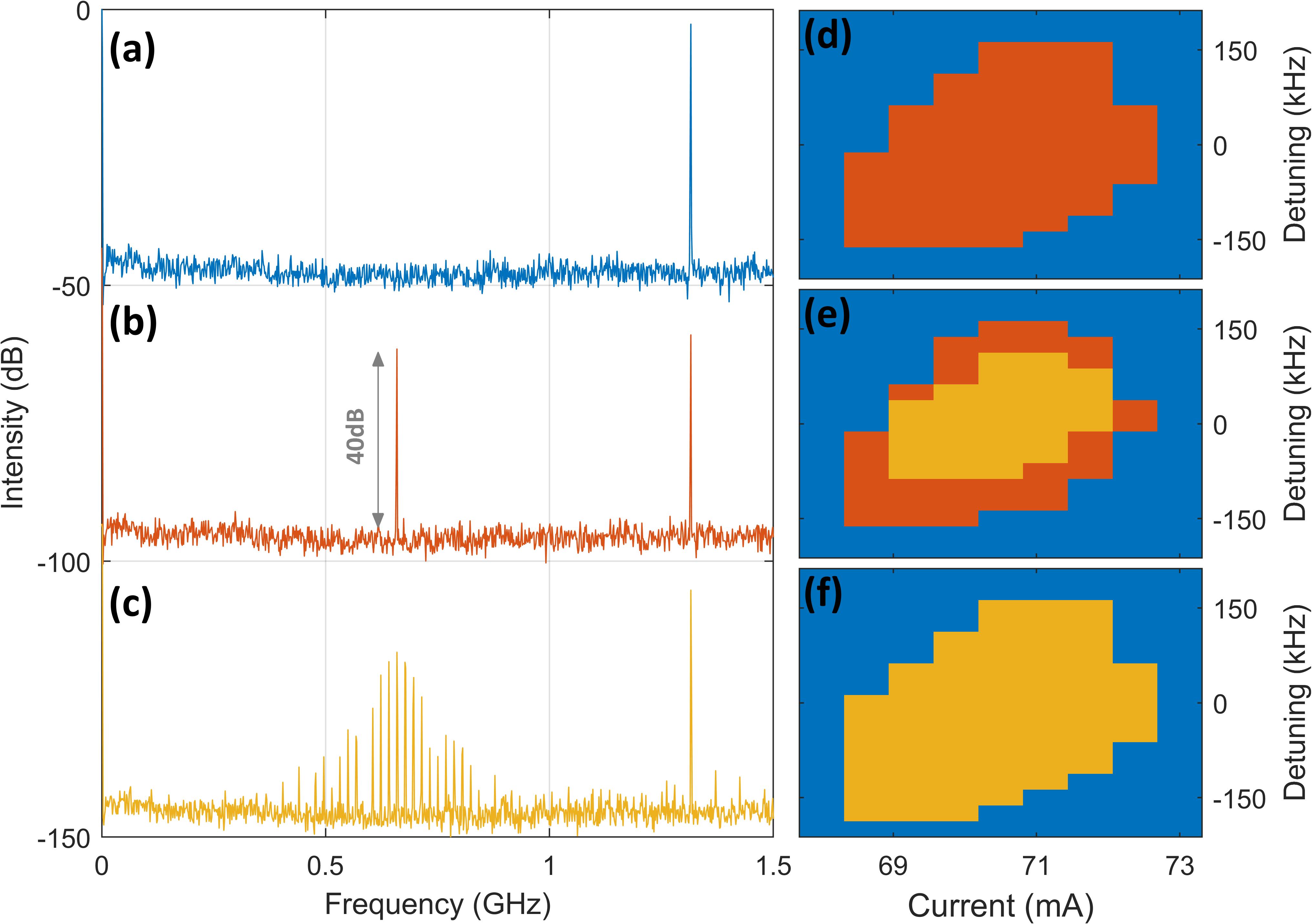} \caption{(a-c) Intensity Fourier spectra corresponding to Fig.~\ref{fig:2}(a-c).
(a) Regular HML regime with $N=72$ pulses whose first peak frequency
aligns with the resonance $\nu=1320.8\,$MHz. (b) Pure TC$^{\pm}$
state with one pulse every two cycles, resulting in lines separated
by half of the modulation frequency. (c) Coexistence between TC$^{+}$
and TC$^{-}$ states. (d-f) Mapping of the dynamical regimes in (a-c)
across gain pumping current and frequency detuning under different
scanning protocols. The colors blue, orange, and yellow correspond
to the regimes observed in (a-c), respectively. (d) Diagram obtained
by adiabatic tuning of the current and modulation frequency in which
only HML and pure TC$^{\pm}$ states are observed. (e) Diagram obtained
by resetting the current below the lasing threshold at each parameter
value. (f) Diagram obtained by adiabatic tuning of the current and
modulation frequency for the nearby odd resonance ($N=71$ pulses
and $\nu=1302.5$).}
\label{fig:3}
\end{figure}

Figures~\ref{fig:3}(a-c) show the low-frequency portion of the Fourier
spectrum intensity time traces corresponding to the regimes depicted
in Fig.~\ref{fig:2}. We observe in Fig.~\ref{fig:3}(a) that, for
the regular HML regime presented in Fig.~\ref{fig:2}(a), the associated
RF frequency comb lines are separated by the modulation frequency.
The figures~\ref{fig:3}(a-c) focus upon the fundamental tone imposed
by the RF modulator showing at least 40~dB of suppression over the
noise floor without active stabilization. However, for a TC$^{+}$
regime depicted in Fig.~\ref{fig:2}(b), the line separation has
been reduced by half as the period of the pulse train is doubled,
see Fig.~\ref{fig:3}(b). Finally, Fig.~\ref{fig:3}(c) shows the
intensity spectrum of the TC$^{+}$/ TC$^{-}$ coexistence state,
where multiple lines appear near half of the modulation frequency
due to presence of defects. These multiple peaks are separated by
a frequency $1/\tau$. This can be understood from the Wiener--Khinchin
theorem that states that the RF spectrum is the Fourier transform
of the intensity autocorrelation. In this case, the presence of defects
create a slow modulation at the round-trip frequency in the autocorrelation.
While pure TC$^{\pm}$ can be maintained indefinitely, mixed TC$^{+}$/TC$^{-}$
states are metastable and relax toward a pure TC$^{\pm}$ state after
a few minutes with $N=72$. 

\begin{figure}
\centering{}\includegraphics[viewport=20bp 0bp 540bp 400bp,clip,width=1\columnwidth]{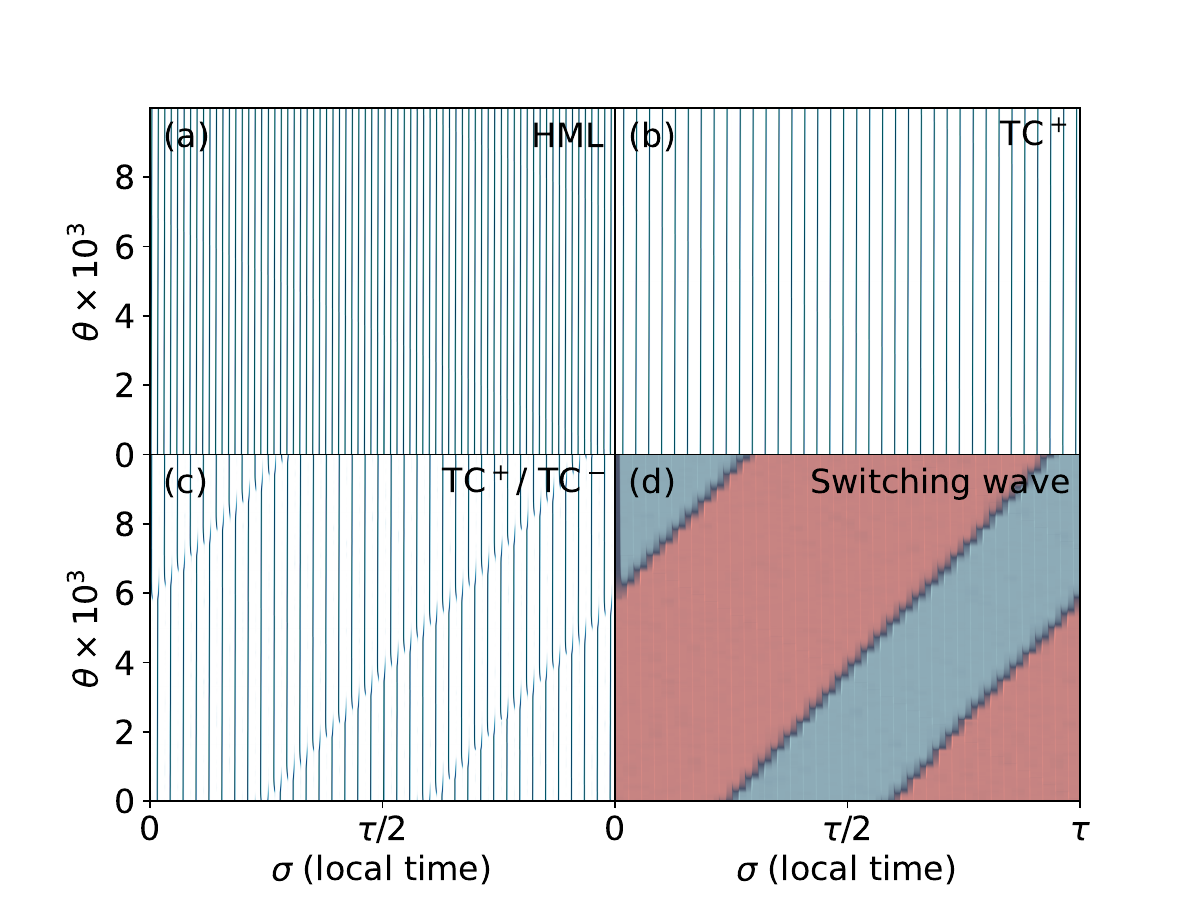}
\caption{Numerical solutions of Eqs.~\eqref{eq:AML1}-\eqref{eq:AML2} in
a bi-dimensional representation obtained for different modulation
frequencies. (a) HML$_{N}$ solution with $N=70$ pulses for $\omega_{m}$$=4.39745$
so that each slot created by the potential is occupied. b) A time-crystal
solution, arising for $\omega_{m}=4.39753$, with every second slot
occupied so that $N/2$ pulses circulate within the cavity. c) Coexistence
between the TC$^{+}$ and TC$^{-}$ states, separated by traveling
defects. d) Processed data showing the dynamics as a switching wave
between the two TC$\pm$ states. Other parameters are: $\gamma=40,\,\kappa_{0}=0.1735,\alpha=1,\,G_{0}=1.44,\,\Gamma=1,\,m=0.42,\,\varphi=-\pi/2,\,\tau=100$.}
\label{fig:4}
\end{figure}

We mapped the observed regimes over the range of the pumping current
and frequency of the RF generator and the resulting phase diagrams
are depicted in Fig.~\ref{fig:3}(d-f). Here, blue, orange and yellow
patches correspond to the HML, pure TC$^{\pm}$ regimes and mixed
TC$^{+}$/TC$^{-}$ states, respectively. One clearly observes the
appearance of an island of stable TC$^{\pm}$ region in Fig.~\ref{fig:3}(d).
To obtain a metastable coexisting TC$^{+}$/TC$^{-}$ state, we modified
the current scanning protocol. Instead of slowly scanning the current
and the modulator frequency, the former is firstly set just below
the lasing threshold before being reset back to any given value. This
approach allows us to restart the system from a noisy initial state
and to generate with a high probability a metastable TC$^{+}$/TC$^{-}$
superposition, as represented in Fig.~\ref{fig:3}(e). Finally, setting
the system around the odd resonance tongue corresponding to $N=71$
pulses, the same adiabatic scanning protocol used in Fig.~\ref{fig:3}(d)
is applied, leading to Fig.~\ref{fig:3}(f). Here, however, only
TC$^{+}$/TC$^{-}$ mixed states composed by an odd number of domain
walls could be observed, in addition to the HML state. We further
noted that the dynamics always contained at least one defect and that
the latter persisted indefinitely. 

Our theoretical model follows the one developed in \citep{KGJ-OL-24}
and in which we assume that the filter bandwidth is much smaller than
the gain broadening. The dynamical model for the field envelope $E$
at the filter output and the population inversion field $G$ read:
\begin{align}
\frac{\dot{E}}{\gamma}+E & =\sqrt{\kappa(t)}e^{\left(1-i\alpha\right)G\left(t-\tau\right)/2}E\left(t-\tau\right)\,,\label{eq:AML1}\\
\dot{G} & =\Gamma\left(G_{0}-G\right)-\left(e^{G}-1\right)\left|E\right|^{2}.\label{eq:AML2}
\end{align}
The round-trip time is $\tau$ and $\left(G_{0},\Gamma\right)$ are
the pumping and gain recovery rates, respectively, while $\alpha$
is the linewidth enhancement factor and $\gamma$ corresponds to the
bandwidth of the spectral filter. The fraction of light intensity
kept in the cavity at the output coupler is $T_{o}$ and the time-dependent
intensity modulator transmission function is $T_{i}(t)$, leading
to $\kappa(t)=T_{o}T_{i}(t)$. While our approach is not limited in
the shape nor in the amplitude of the intensity modulation, we choose
that corresponding to a MZM modulator $T_{i}(t)=\left\{ 1+\cos\left[\varphi+2m\cos\left(\omega_{m}t\right)\right]\right\} /2,$
where $\omega_{m}$ is the modulation frequency and $m$ is the modulation
depth. We define $\kappa_{0}=T_{o}/2$ which corresponds to the average
cavity transmission when the modulator is operated in quadrature ($\varphi=\pm\pi$).
We operate the AML system given by \eqref{eq:AML1},\eqref{eq:AML2}
with a modulator operating in its linear transmission regime for which
$\varphi=-\pi/2$ and at a frequency $\omega_{m}$ at which the modulator
creates a potential with $N=70$ slots, \emph{i.e.} $\omega_{m}\simeq\left(2\pi/\tau\right)N$.
Increasing the pumping current $G_{0}$ slightly above the lasing
threshold leads to the formation of the HML solution with a pulse
located at each minima of the potential, see Fig.~\ref{fig:4}(a).
Here, a bi-dimensional representation of the last $10^{4}$ round-trips
(from a total of $5\times10^{5}$) is presented. As each slot created
by the potential is occupied, the period of the resulting HML solution
is equal to that of the modulation. However, for a slight blue shift
of the modulation frequency $\omega_{m}$, the HML state becomes unstable,
instead promoting a solution at twice the period of the modulation,
where pulses occupy only the even (TC$^{+}$) or odd (TC$^{-}$) positions,
forming discrete TC states. This is shown in Fig.~\ref{fig:4}(b). 

\begin{figure}[b]
\centering \includegraphics[width=0.9\columnwidth]{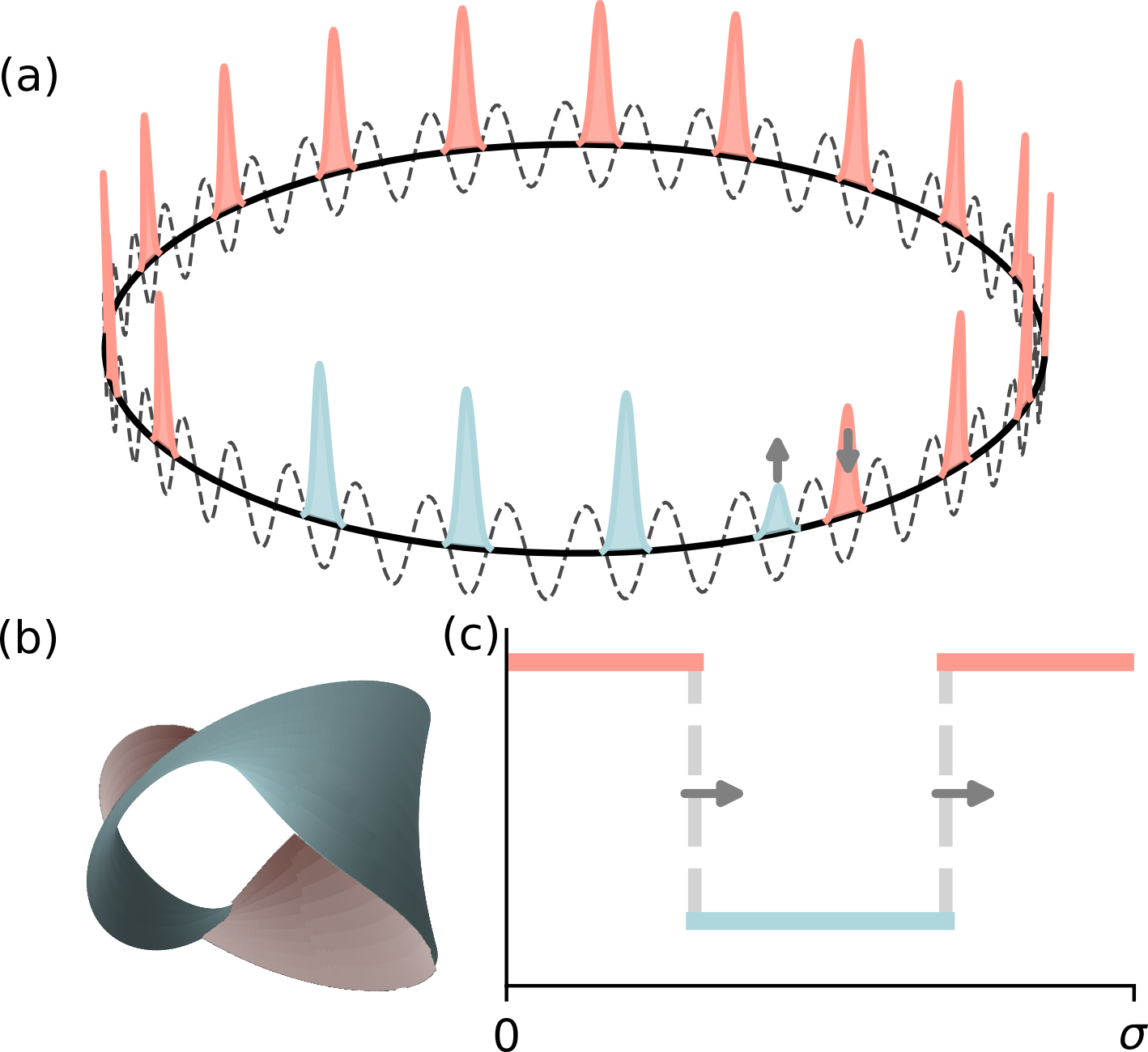} \caption{a) Sketch of the dynamics when the system is in the superposed TC$^{+}$/TC$^{-}$
configuration allowing for coexisting crystallites of TC$^{+}$ (pink)
and T$C^{-}$ (blue) in the cavity. Arrows indicate the motion of
the of defects in the boundary regions. b) Representation of (a) as
a twisted Möbius strip closed onto itself. c) Domain wall transitions
between the two states traveling at equal speed in the absence of
noise.}
\label{fig:5}
\end{figure}

In agreement with the experimental findings, we also observed that
the two TC$^{\pm}$ states may coexist for long durations. Hence,
in the course of self-organization from the off solution in the presence
of noise, the laser may form many small domains (or crystallites)
of the two kind of equivalent bistable TC$^{\pm}$ states with different
sizes. The latter are separated by defects, or domain walls, see Fig.~\ref{fig:4}(c).
The resulting state is schematically depicted in Fig.~\ref{fig:5}(a)
where the arrows indicate how the defects travel around the cavity.
As the time simulation presented in Fig.~\ref{fig:4}(c) was performed
in the presence of a Gaussian white noise of the amplitude $2\times10^{-3}$,
we observe that the created defects move relative to each other due
to noise leading to pairwise annihilation and coarsening. This leads
to a situation with in which more distant domain walls travel at equal
speed and only weakly interact with each other as their relative distance
is almost constant over the duration of very long time simulations,
which explains the experimentally observed longevity of the defect
pairs, \emph{cf.} Fig.~\ref{fig:2}(c), whose lifetime is inversely
proportional to the amplitude of the system noise. Note, that for
an even number of slots in the potential, only even numbers of defects
can exist. Even though we observed, experimentally and theoretically,
situations with more than two defects, they always appear in pairs.
This can be understood interpreting the transitions as twists of a
two-colored Möbius strip, where the pink (blue) sides correspond to
the TC$^{+}$ (TC$^{-}$) states, respectively. If a section of the
band is flipped, this always creates a pair of twists, \emph{cf.}
Fig.~\ref{fig:5}(b). Continuing this analogy, the case of an ``odd''
number of potential slots as discussing for $N=71$ in Fig.~\ref{fig:3}(f)
leads to a period-two solution that does not fit the domain, automatically
creating an odd number of defects. Finally, in agreement with the
experimental Fig.~\ref{fig:2}(d), the defects can be interpreted
as domain walls between the two bistable TC$^{\pm}$ states ($y_{\pm}=\pm1$)
as schematically shown in Fig.~\ref{fig:5}(c). We post-processed
the simulation data from Fig.~\ref{fig:4}(c) as in the experiment
leading to an excellent agreement and a similar phenomenology, stressing
again the domain walls analogy between two TC states, see Fig.~\ref{fig:4}(d).

In conclusion, we have provided the first experimental evidence of
spontaneously emerging time-crystal states in an actively mode-locked
semiconductor laser. These states are robust and persist indefinitely,
exhibit bistability and domain-wall dynamics between equivalent configurations,
in full agreement with a time-delayed theoretical model tailored for
large gain and loss. Beyond their fundamental significance in breaking
discrete time-translation symmetry in a many-body photonic system,
our results establish AML semiconductor lasers as a versatile, accessible,
and highly controllable platform for exploring non-equilibrium phases
of light opening pathways to harness time-crystal behavior for precision
timing and frequency-comb control. Further works shall consider breaking
the symmetry between the two equivalent TC states which should allow
for a more precise control of the dynamics as well as the exploration
of TC states of higher order as depicted in Appendix B.
\begin{acknowledgments}
We acknowledge the financial support of the project KOGIT, Agence
Nationale de la Recherche (No. ANR-22-CE92-0009) and Deutsche Forschungsgemeinschaft
(DFG) via Grant No. 505936983. RW, JYS and JJ acknowledge funding
from Ministerio de Economía y Competitividad (PID2021-128910NB-100
AEI/FEDER UE, PGC2018-099637-B-100 AEI/FEDER UE). 
\end{acknowledgments}

%\bibliographystyle{unsrt}
%\bibliography{full_140723,AML_bib,extra,extra2}

\begin{thebibliography}{10}

\bibitem{W-PRL-12}
Frank Wilczek.
\newblock Quantum time crystals.
\newblock {\em Phys. Rev. Lett.}, 109:160401, Oct 2012.

\bibitem{B-PRL-13}
Patrick Bruno.
\newblock Impossibility of spontaneously rotating time crystals: A no-go
  theorem.
\newblock {\em Phys. Rev. Lett.}, 111:070402, Aug 2013.

\bibitem{WO-PRL-15}
Haruki Watanabe and Masaki Oshikawa.
\newblock Absence of quantum time crystals.
\newblock {\em Phys. Rev. Lett.}, 114:251603, Jun 2015.

\bibitem{S-PRA-15}
Krzysztof Sacha.
\newblock Modeling spontaneous breaking of time-translation symmetry.
\newblock {\em Phys. Rev. A}, 91:033617, Mar 2015.

\bibitem{EBN-PRL-16}
Dominic~V. Else, Bela Bauer, and Chetan Nayak.
\newblock Floquet time crystals.
\newblock {\em Phys. Rev. Lett.}, 117:090402, Aug 2016.

\bibitem{KLM-PRL-16}
Vedika Khemani, Achilleas Lazarides, Roderich Moessner, and S.~L. Sondhi.
\newblock Phase structure of driven quantum systems.
\newblock {\em Phys. Rev. Lett.}, 116:250401, Jun 2016.

\bibitem{GHU-PRL-18}
Zongping Gong, Ryusuke Hamazaki, and Masahito Ueda.
\newblock Discrete time-crystalline order in cavity and circuit qed systems.
\newblock {\em Phys. Rev. Lett.}, 120:040404, Jan 2018.

\bibitem{BTJ-NC-19}
Berislav Bu{\v{c}}a, Joseph Tindall, and Dieter Jaksch.
\newblock Non-stationary coherent quantum many-body dynamics through
  dissipation.
\newblock {\em Nat. Commun.}, 10(1):1730, 2019.

\bibitem{SZ-RPP-18}
Krzysztof Sacha and Jakub Zakrzewski.
\newblock Time crystals: a review.
\newblock {\em Reports on Progress in Physics}, 81(1):016401, nov 2017.

\bibitem{S-TC-2020}
K.~Sacha.
\newblock {\em Time Crystals}.
\newblock Springer Series on Atomic, Optical, and Plasma Physics. Springer
  International Publishing, 2020.

\bibitem{YN-PT-18}
Norman~Y. Yao and Chetan Nayak.
\newblock Time crystals in periodically driven systems.
\newblock {\em Physics Today}, 71(9):40--47, 09 2018.

\bibitem{ZHK-N-17}
Jiehang Zhang, Paul~W Hess, A~Kyprianidis, Petra Becker, A~Lee, J~Smith,
  Gaetano Pagano, I-D Potirniche, Andrew~C Potter, Ashvin Vishwanath, et~al.
\newblock Observation of a discrete time crystal.
\newblock {\em Nature}, 543(7644):217--220, 2017.

\bibitem{CCL-N-17}
Soonwon Choi, Joonhee Choi, Renate Landig, Georg Kucsko, Hengyun Zhou, Junichi
  Isoya, Fedor Jelezko, Shinobu Onoda, Hitoshi Sumiya, Vedika Khemani, et~al.
\newblock Observation of discrete time-crystalline order in a disordered
  dipolar many-body system.
\newblock {\em Nature}, 543(7644):221--225, 2017.

\bibitem{FR-SA-22}
Philipp Frey and Stephan Rachel.
\newblock Realization of a discrete time crystal on 57 qubits of a quantum
  computer.
\newblock {\em Sci. Adv.}, 8(9):eabm7652, 2022.

\bibitem{MIQ-N-22}
Xiao Mi, Matteo Ippoliti, Chris Quintana, Ami Greene, Zijun Chen, Jonathan
  Gross, Frank Arute, Kunal Arya, Juan Atalaya, Ryan Babbush, et~al.
\newblock Time-crystalline eigenstate order on a quantum processor.
\newblock {\em Nature}, 601(7894):531--536, 2022.

\bibitem{KKP-PRL-21}
Hans Ke\ss{}ler, Phatthamon Kongkhambut, Christoph Georges, Ludwig Mathey,
  Jayson~G. Cosme, and Andreas Hemmerich.
\newblock Observation of a dissipative time crystal.
\newblock {\em Phys. Rev. Lett.}, 127:043602, Jul 2021.

\bibitem{TMM-NAC-22}
Hossein Taheri, Andrey~B. Matsko, Lute Maleki, and Krzysztof Sacha.
\newblock All-optical dissipative discrete time crystals.
\newblock {\em Nature Communications}, 13(1):848, Feb 2022.

\bibitem{TAM-SR-22}
Jordi Tiana-Alsina and Cristina Masoller.
\newblock Time crystal dynamics in a weakly modulated stochastic time delayed
  system.
\newblock {\em Scientific Reports}, 12(1):4914, Mar 2022.

\bibitem{KGJ-OL-24}
Elias~R. Koch, Svetlana~V. Gurevich, and Julien Javaloyes.
\newblock Pulse instabilities in harmonic active mode-locking: a time-delayed
  approach.
\newblock {\em Opt. Lett.}, 49(19):5663--5666, Oct 2024.

\bibitem{HFP-APL-64}
L.~E. Hargrove, R.~L. Fork, and M.~A. Pollack.
\newblock Locking of he-ne laser modes induced by synchronous intracavity
  modulation.
\newblock {\em Applied Physics Letters}, 5(1):4--5, 1964.

\bibitem{WSY_APLP_16}
Yu~Wang, Sze~Y Set, and Shinji Yamashita.
\newblock Active mode-locking via pump modulation in a tm-doped fiber laser.
\newblock {\em APL Photonics}, 1(7), 2016.

\bibitem{HMCD_JLT_00}
M.~Horowitz, C.R. Menyuk, T.F. Carruthers, and I.N. Duling.
\newblock Theoretical and experimental study of harmonically modelocked fiber
  lasers for optical communication systems.
\newblock {\em J. of Lightwave Technol.}, 18(11):1565--1574, 2000.

\bibitem{HZU_IEE_01}
R~Holzwarth, M~Zimmermann, Thomas Udem, and TW~Hansch.
\newblock Optical clockworks and the measurement of laser frequencies with a
  mode-locked frequency comb.
\newblock {\em IEEE J. of Quantum Electronics}, 37(12):1493--1501, 2001.

\bibitem{HUH_PRL_00}
R~Holzwarth, Th~Udem, Th~W H{\"a}nsch, JC~Knight, WJ~Wadsworth, and P~St~J
  Russell.
\newblock Optical frequency synthesizer for precision spectroscopy.
\newblock {\em Phys. Rev. Lett.}, 85(11):2264, 2000.

\bibitem{UHH_N_02}
Th~Udem, Ronald Holzwarth, and Theodor~W H{\"a}nsch.
\newblock Optical frequency metrology.
\newblock {\em Nature}, 416(6877):233--237, 2002.

\bibitem{K_Nt_03}
Ursula Keller.
\newblock Recent developments in compact ultrafast lasers.
\newblock {\em Nature}, 424(6950):831--838, 2003.

\bibitem{AJ-BOOK-17}
E.~Avrutin and J.~Javaloyes.
\newblock {\em Mode-Locked Semiconductor Lasers, Book Chapter In: Handbook of
  Optoelectronic Device Modeling and Simulation}.
\newblock CRC press, Taylor and Francis, United Kingdom, 2017.

\bibitem{ST_OLT_15}
L.A.M. Saito and E.A. {Thoroh de Souza}.
\newblock Identifying the mechanisms of pulse formation and evolution in
  actively mode-locked erbium fiber lasers with meters and kilometers-long.
\newblock {\em Opt. Laser Technol.}, 71:16--21, 2015.

\bibitem{RHWC_NC_16}
DG~Revin, M~Hemingway, Y~Wang, JW~Cockburn, and A~Belyanin.
\newblock Active mode locking of quantum cascade lasers in an external ring
  cavity.
\newblock {\em Nat. Commun.}, 7(1):11440, 2016.

\bibitem{YZC_OE_20}
Bo~Yang, Hongyan Zhao, Zizheng Cao, Shuna Yang, Yanrong Zhai, Jun Ou, and Hao
  Chi.
\newblock Active mode-locking optoelectronic oscillator.
\newblock {\em Opt. Express}, 28(22):33220--33227, 2020.

\bibitem{TK_IEE_09}
Jonathan Tu and Jose~Nathan Kutz.
\newblock Pulse formation, harmonic mode-locking, and stability in actively
  mode-locked laser cavities.
\newblock {\em IEEE J. of Quantum Electronics}, 45(3):282--291, 2009.

\bibitem{CDDC_OL_09}
C.~Cuadrado-Laborde, A.~Diez, M.~Delgado-Pinar, J.~L. Cruz, and M.~V.
  Andr\'{e}s.
\newblock Mode locking of an all-fiber laser by acousto-optic superlattice
  modulation.
\newblock {\em Opt. Lett.}, 34(7):1111--1113, 2009.

\bibitem{KKL_PhRes_17}
Jihwan Kim, Joonhoi Koo, and Ju~Han Lee.
\newblock All-fiber acousto-optic modulator based on a cladding-etched optical
  fiber for active mode-locking.
\newblock {\em Photon. Res.}, 5(5):391--395, 2017.

\bibitem{KKWK_OL_94}
D.~Kopf, F.~X. K\"{a}rtner, K.~J. Weingarten, and U.~Keller.
\newblock Pulse shortening in a nd:glass laser by gain reshaping and soliton
  formation.
\newblock {\em Opt. Lett.}, 19(24):2146--2148, 1994.

\bibitem{PYZ_OC_02}
Can Peng, Minyu Yao, Janfeng Zhang, Hongming Zhang, Qianfan Xu, and Yizhi Gao.
\newblock Theoretical analysis of actively mode-locked fiber ring laser with
  semiconductor optical amplifier.
\newblock {\em Optics Communications}, 209(1):181--192, 2002.

\bibitem{ZHM_OC_05}
K.E. Zoiros, T.~Houbavlis, and M.~Moyssidis.
\newblock Complete theoretical analysis of actively mode-locked fiber ring
  laser with external optical modulation of a semiconductor optical amplifier.
\newblock {\em Optics Communications}, 254(4):310--329, 2005.

\bibitem{PGG-NAC-20}
A.~M. Perego, B.~Garbin, F.~Gustave, S.~Barland, F.~Prati, and G.~J.
  de~Valc\'{a}rcel.
\newblock Coherent master equation for laser modelocking.
\newblock {\em Nat. Commun.}, 11(1):311, 2020.

\bibitem{BWX_AFM_18}
Jakub Bogus{\l}awski, Yadong Wang, Hui Xue, Xiaoxia Yang, Dong Mao, Xuetao Gan,
  Zhaoyu Ren, Jianlin Zhao, Qing Dai, Grzegorz Sobo{\'n}, et~al.
\newblock Graphene actively mode-locked lasers.
\newblock {\em Advanced Functional Materials}, 28(28):1801539, 2018.

\bibitem{ZPH-JQE-05}
A.~Zeitouny, Y.N. Parkhomenko, and M.~Horowitz.
\newblock Stable operating region in a harmonically actively mode-locked fiber
  laser.
\newblock {\em IEEE Journal of Quantum Electronics}, 41(11):1380--1387, 2005.

\bibitem{ZH-JLT-06}
Zeitouny and Horowitz.
\newblock Experimental study of pulse recovery from dropout in an actively
  mode-locked fiber laser.
\newblock {\em Journal of Lightwave Technology}, 24(10):3671--3676, 2006.

\bibitem{LSB-OECC-07}
Huy~Quoc Lam, P.~Shum, Le~Nguyen Binh, Y.D. Gong, Ming Tang, and Songnian Fu.
\newblock Pulse dropout and subharmonic locking in active mode-locked
  birefringence fiber laser.
\newblock In {\em Proceedings of the 2007 Opto-Electronics and Communications
  Conference (OECC-IOOC)}, 2007.

\bibitem{YKN-APL-92}
Eiji Yoshida, Yasuo Kimura, and Masataka Nakazawa.
\newblock Laser diode‐pumped femtosecond erbium‐doped fiber laser with a
  sub‐ring cavity for repetition rate control.
\newblock {\em Applied Physics Letters}, 60(8):932--934, 02 1992.

\bibitem{HM-OL-93}
G.~T. Harvey and L.~F. Mollenauer.
\newblock Harmonically mode-locked fiber ring laser with an internal
  fabry--perot stabilizer for soliton transmission.
\newblock {\em Opt. Lett.}, 18(2):107--109, Jan 1993.

\bibitem{SS-EL-93}
X.~Shan and D.M. Spirit.
\newblock Novel method to suppress noise in harmonically modelocked erbium
  fibre lasers.
\newblock {\em Electronics Letters}, 29:979--981, 1993.

\bibitem{DHI-OL-94}
C.~R. Doerr, H.~A. Haus, E.~P. Ippen, M.~Shirasaki, and K.~Tamura.
\newblock Additive-pulse limiting.
\newblock {\em Opt. Lett.}, 19(1):31--33, Jan 1994.

\bibitem{CD-OL-96}
Thomas~F. Carruthers and Irl~N. Duling.
\newblock 10-ghz, 1.3-ps erbium fiber laser employing soliton pulse shortening.
\newblock {\em Opt. Lett.}, 21(23):1927--1929, Dec 1996.

\bibitem{TGK-OL-00}
E.~R. Thoen, M.~E. Grein, E.~M. Koontz, E.~P. Ippen, H.~A. Haus, and L.~A.
  Kolodziejski.
\newblock Stabilization of an active harmonically mode-locked fiber laser using
  two-photon absorption.
\newblock {\em Opt. Lett.}, 25(13):948--950, Jul 2000.

\bibitem{KCB-JQE-98}
J.N. Kutz, B.C. Collings, K.~Bergman, and W.H. Knox.
\newblock Stabilized pulse spacing in soliton lasers due to gain depletion and
  recovery.
\newblock {\em Quantum Electronics, IEEE Journal of}, 34(9):1749--1757, Sep
  1998.

\bibitem{CJM-PRA-16}
P.~Camelin, J.~Javaloyes, M.~Marconi, and M.~Giudici.
\newblock Electrical addressing and temporal tweezing of localized pulses in
  passively-mode-locked semiconductor lasers.
\newblock {\em Phys. Rev. A}, 94:063854, Dec 2016.

\bibitem{AGL-PRA-92}
F.~T. Arecchi, G.~Giacomelli, A.~Lapucci, and R.~Meucci.
\newblock Two-dimensional representation of a delayed dynamical system.
\newblock {\em Phys. Rev. A}, 45:R4225--R4228, Apr 1992.

\bibitem{GMZ-EPL-12}
Giovanni Giacomelli, Francesco Marino, Michael~A. Zaks, and Serhiy Yanchuk.
\newblock Coarsening in a bistable system with long-delayed feedback.
\newblock {\em EPL (Europhysics Letters)}, 99(5):58005, 2012.

\bibitem{JAH-PRL-15}
J.~Javaloyes, T.~Ackemann, and A.~Hurtado.
\newblock Arrest of domain coarsening via antiperiodic regimes in delay
  systems.
\newblock {\em Phys. Rev. Lett.}, 115:203901, Nov 2015.

\end{thebibliography}

\section*{End Matter}

\paragraph*{Appendix A: Experimental setup:---}

Figure 6 shows the experimental setup, which consists of a Semiconductor
Optical Amplifier (BOA-1004P from Thorlabs), a Mach-Zehnder Intensity
Modulator (MX-LN-10 from Exail) driven by an RF signal generator (Marconi
IFR 2024), an optical isolator (ISO) to ensure unidirectional light
propagation, and a 90/10 optical coupler. The total cavity length
is approximately $L=11\,$m corresponding to a fundamental frequency
of $18.3\,$MHz and a round-trip time of $\tau=54.5\,$ns. At the
coupler, 90\% of the optical power is recirculated inside the cavity,
while 10\% is sent to a $12\,$GHz bandwidth photodetector (Newport 1544-B).
The detected signal is then analyzed using a $12\,$GHz bandwidth
oscilloscope (Agilent Infinium DSO 81204A) and a radio-frequency intensity
spectrum analyzer (Anritsu MS2602A). The optical spectrum is monitored
with an HP~86142A optical spectrum analyzer with $0.06\,$nm resolution.

\paragraph*{Appendix B: Period 3 regimes---}

In addition to the TC$^{\pm}$ regimes, where every other pulse is
emitted, we also observed situations in which one pulse is emitted
every three modulation cycles as shown in Fig.~7. We note that such
configurations do not lead to defects since $N=72$ is divisible 3.

\begin{figure}[H]
\includegraphics[width=1\columnwidth]{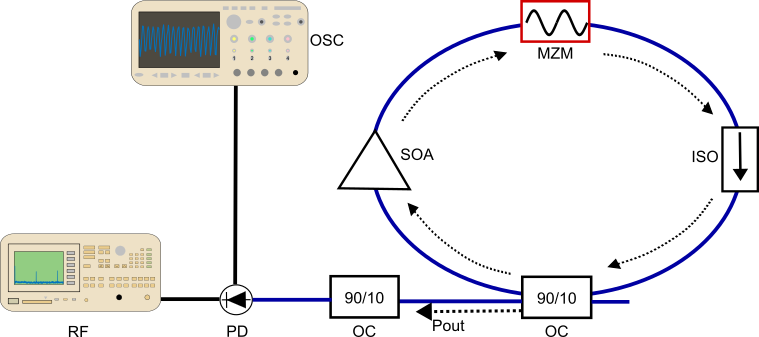}

\caption{Experimental setup of the active mode-locking cavity and detection
system. MZM: Mach-Zehnder Modulator. ISO: Optical Isolator. OC: Optical
Coupler. SOA: Semiconductor Optical Amplifier. PD: Photodetector.
OSC: Oscilloscope. RF: Intensity Spectrum Analyzer. Light propagates
in the cavity along the direction of the dotted arrows.}
\end{figure}

\begin{figure}[H]
\includegraphics[clip,width=1\columnwidth]{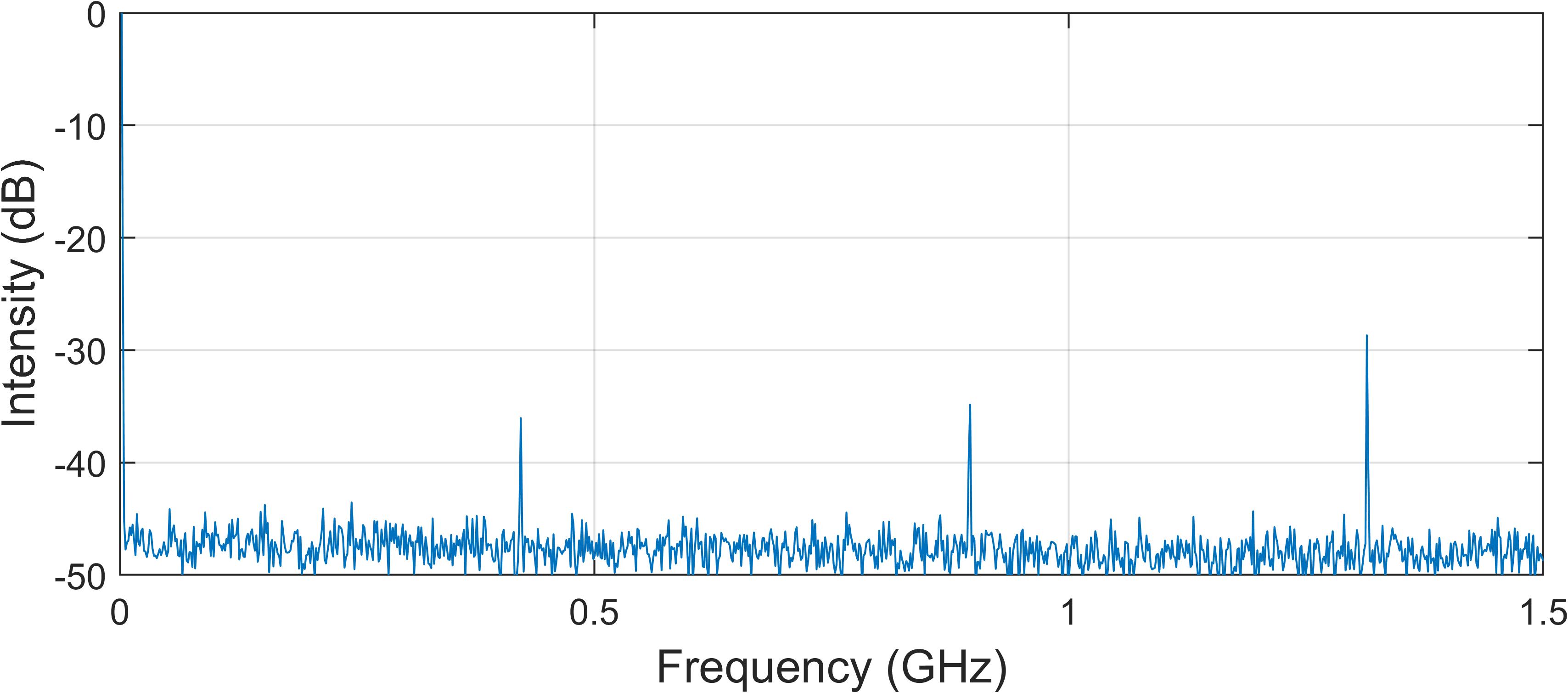}

\caption{Intensity Fourier Spectrum of a TC regime with one pulse every three
cycles, leading to lines separated by one third of the resonance frequency.
The harmonic number is $N=72$ and the parameters correspond to that
of Fig.~3(e).}
\end{figure}

\end{document}